\pdfoutput=1
\documentclass[pra,twocolumn,10pt,aps,superscriptaddress]{revtex4-1}
\usepackage[utf8]{inputenc}
\usepackage{hyperref}
\usepackage{amsmath}
\usepackage{amsfonts}
\usepackage{amssymb}
\usepackage{graphicx}

\usepackage{color}
\definecolor{mygreen}{rgb}{0,0.5,0} 
\definecolor{mygrey}{rgb}{0.5,0.5,0.5} 
\definecolor{myred}{rgb}{0.75,0,0} 
\definecolor{myblue}{rgb}{0,0,0.75} 
\definecolor{mymagenta}{cmyk}{0,1,0,0.12} 
\definecolor{mycyan}{cmyk}{1,0,0,0.12} 
\definecolor{myorange}{rgb}{1,0.5,0}

\usepackage{siunitx}
\renewcommand{\vec}[1]{\mathbf{#1}}

\newcommand{\mean}[1]{\langle #1 \rangle}

\newcommand{\var}[1]{\mathrm{var} \left(  #1 \right) }

\newcommand{\Tzero}{\ensuremath{\mathcal{T}_0} }
\newcommand{\Tone}{\ensuremath{\mathcal{T}_1} }
\newcommand{\Ttwo}{\ensuremath{\mathcal{T}_2} }
\newcommand{\Ttwom}{\ensuremath{\overline{\mathcal{T}_2}} }
\newcommand{\rb}{\ensuremath{ ^{87}\mathrm{Rb} } }
\newcommand{\spinor}[1]{\left( #1 \right)^T }

\newcommand{\transition}[2]{\ensuremath{| #1 \rangle \rightarrow | #2 \rangle}}
\newcommand{\state}[1]{\ensuremath{| #1 \rangle}}
\newcommand{\Ntot}{\ensuremath{N_{\rm tot}}}
\newcommand{\NL}{\ensuremath{N_L}}

\newcommand{\opf}[1]{\ensuremath{\hat{f}_{#1}}}
\newcommand{\opF}[1]{\ensuremath{\hat{F}_{#1}}}
\newcommand{\opj}[1]{\ensuremath{\hat{\jmath}_{#1}}}

\newcommand{\ICFOAddress}{ICFO--Institut de Ciencies Fotoniques, The Barcelona Institute of Science and Technology, 08860 Castelldefels (Barcelona), Spain}
\newcommand{\ICREAAddress}{ICREA--Instituci\'o Catalana de Recerca i Estudis Avan\c{c}ats, 08010 Barcelona, Spain}

\begin{document}
 
%opening
\title{Multi-second magnetic coherence in a single domain spinor Bose-Einstein condensate}

\author{Silvana Palacios }
\affiliation{\ICFOAddress}
\author{Simon Coop}
\affiliation{\ICFOAddress}
\author{Pau Gomez}
\affiliation{\ICFOAddress}
\author{Thomas Vanderbruggen}
\affiliation{Koheron, Centre scientifique d'Orsay, Batiment 503, 91400 Orsay, France}
\author{Y. Natali Martinez de Escobar}
\affiliation{Lone Star College, University Park, Chemistry and Physics Department, Houston, TX 77070}
\author{Martijn Jasperse}
\affiliation{School of Physics, The University of Melbourne, Parkville, VIC 3010, Australia}
\author{Morgan W. Mitchell}
\affiliation{\ICFOAddress}
\affiliation{\ICREAAddress}

\date{\today}

\begin{abstract}

We describe a compact, robust and versatile system for studying magnetic dynamics in a spinor Bose-Einstein condensate.  Condensates of \rb are produced by all-optical evaporation in a \SI{1560}{\nano\meter} optical dipole trap, using a non-standard loading sequence that employs an auxiliary \SI{1529}{\nano\meter} beam for partial compensation of the strong differential light shift induced by the dipole trap itself.  We use near-resonance Faraday rotation probing to non-destructively track the condensate magnetization, and demonstrate few-Larmor-cycle tracking with no detectable degradation of the spin polarization.  In the ferromagnetic $F=1$ ground state, we observe magnetic $\Tone$ and $\Ttwo^*$ coherence times limited only by the several-second residence time of the atoms in the trap.
\end{abstract}
\maketitle

\section{Introduction}
Spinor BECs are interesting and useful for several fundamental and practical reasons. The rich nature of a spinor condensate has allowed the exploration of different quantum phase transitions and  spontaneous symmetry breaking \cite{Sadler2006},  spin-wave formation \cite{Gu2004},   spin textures \cite{Eto2014},  and a variety of topological excitations such as vortices,  skyrmions \cite{ChoiPRL2012} and Dirac monopoles \citep{Pietila2009,Ray2014}. 
 
 It has been proposed to use this multicomponent system to form  Schr\"odinger cat states  \citep{Cirac1998, Higbie2004} and observe the Einstein-de Haas effect in both strong \cite{Kawaguchi2006edh} and weak dipolar  species \cite{Gawryluk2007}. 
Relevant to the study of  atom laser physics, spinor BECs have been proposed as a way to observe  suppression of quantum phase diffusion \cite{Law1998}. In the field of metrology and quantum information, spinor condensate systems have proven very effective at generating metrologically-useful entanglement and spin squeezing  \cite{Duan2002, Mustecapl2002}. Spinor condensates can be exploited for field sensing, since the commonly-used model of quantum-limited sensitivity  \citep{Budker2007} situates them in the most promising regime due to their high spatial resolution and inherent long  temporal coherence.  2D spatially resolved magnetometers with performance close to projection noise have being demonstrated \cite{Vengalattore2007}, and the possibility to go beyond this standard quantum limit has being studied \citep{Muessel2014, Brask2015}. 

 Here we report on the realization of a spinor condendansate in the $F=1$ ferromagnetic manifold of $\rb$ which occupies a single spatial spin domain. In this regime the single-mode approximation (SMA) can accurately describe the physics of the system \cite{Law1998, koashi2000, Yi2002, Corre2015}. Using nondestructive Faraday rotation probing, we demonstrate a magnetic coherence time of at least several seconds. We observe a dynamics where the quadratic Zeeman modulates the Larmor precession without dephasing, in contrast to similar experiments where  the system could break into different domains causing decoherence \cite{kronjager2005}. In our system  only atom losses degrade the macroscopic spin state. These losses are dominated by one-body losses because of the low densities of our system, in contrast to other experiments where  three-body losses represent the main loss mechanism  \cite{Burt1997, Soding1999, miesner1999, schmaljohann2004}.
 
This work is organized as follows:  \autoref{sec:app} describes our experimental approach to form a spinor condensate in an arbitrary spin state. It is  comprised of a minimalist design of the apparatus an all-optical evaporation in an optical dipole trap (ODT).  In this section we briefly describe our loading technique which exploits the large differential light shift exerted by the ODT to create an effective dark-MOT. 

In \autoref{sec:spat} we support the claim that the size and density of the condensate put it within the SMA regime. In \autoref{sec:oscB} we describe the Zeeman dynamics of an atomic ensemble  in the presence of  a magnetic field. In \autoref{sec:farprob} we describe the non-destructive Faraday rotation measurement implementation and characterization, which we employ to read out the spin state of the atoms. Finally in  \autoref{sec:relaxation} we show the spinor condensate is  immune to most decoherence mechanisms which allows the spin state to remain coherent on the scale of seconds. 

\section{Apparatus and state preparation \label{sec:app}}
\begin{figure}[ht]
  \includegraphics[trim=1cm 0cm 1cm 0cm, clip=true,  width=\columnwidth]{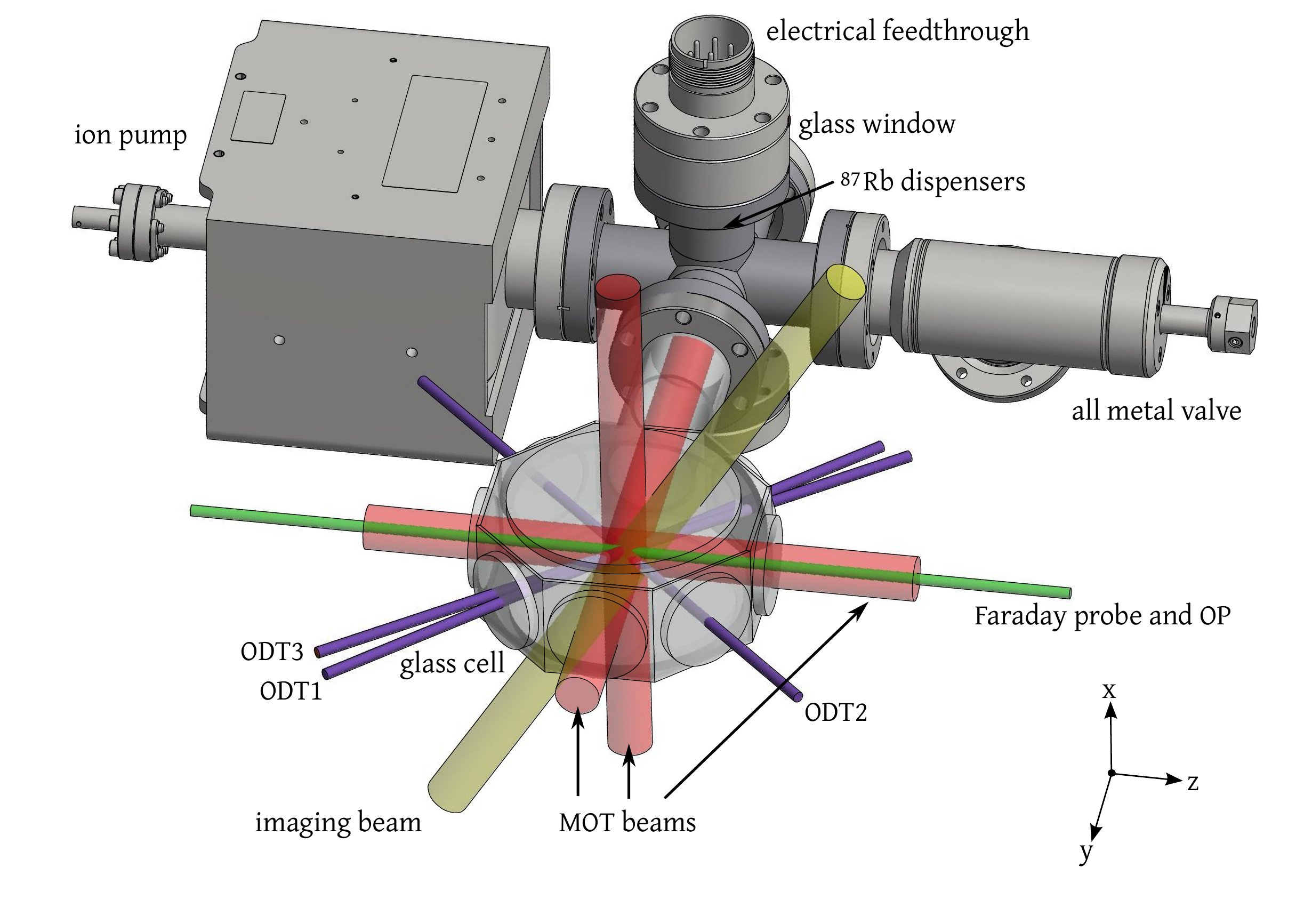}
  \includegraphics[trim=0cm 4.5cm 0cm 0cm, clip=true,  width=\columnwidth]{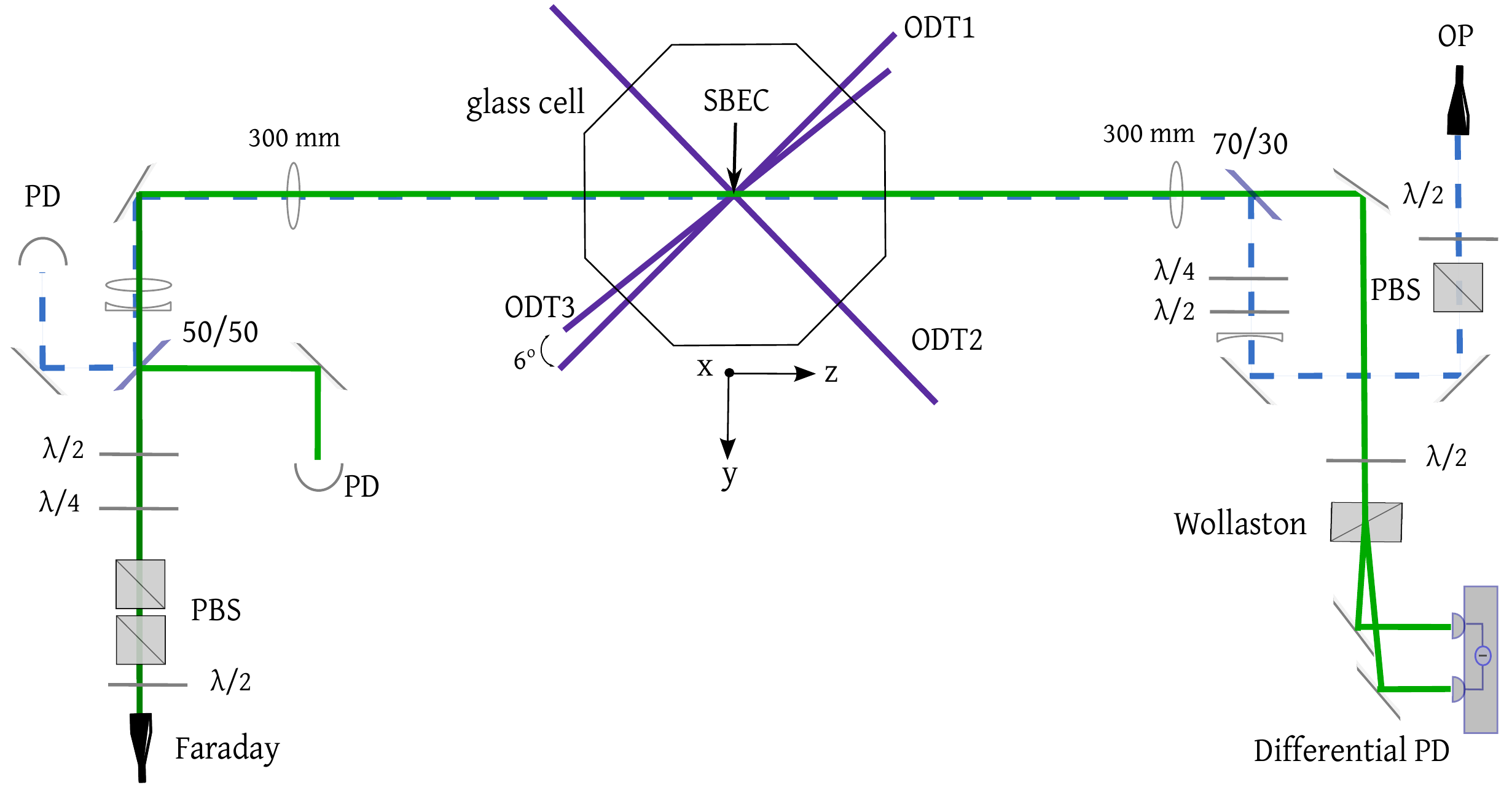}
 \caption{General setup  of the experiment to form a spinor condensate in a glass cell (above). Detailed setup to perform Faraday rotation measurements of the spin state.  Continuous green lines represent the Faraday beam and  dashed blue lines the optical pumping beam (below).  }
 \label{fig:setup}
\end{figure}

As shown  in \autoref{fig:setup}, the vacuum system consists of an all-glass, 9-window enclosure (Octagonal BEC Cell 4, Precision Glassblowing) in which an ultimate pressure of \SI{e-11}{Torr} can be maintained  with  a single pumping element  (TiTan 25SVW, Gamma Vacuum).  The glass cell is AR coated for \SI{780}{nm} and \SI{1560}{nm} to reach  single-window transmission of 97\% and 99\% respectively. The ion pump is shielded with a  high-permeability enclosure which reduces the magnetic field produced by its  magnets by a factor of $\sim400$, such that the field around the glass cell is mainly due to the earth's magnetic field.  \rb is deposited in the chamber by sublimating rubidium from dispensers  mounted inside (Alvasource-3-Rb87-C, Alvatec).  Following activation of the dispensers the pressure rises to about \SI{2e-10}{Torr}, which is the typical pressure of the experiment. 

\begin{figure}[ht]
\includegraphics[trim=0cm 2cm 12.5cm 0cm, clip=true,  width=0.5 \columnwidth]{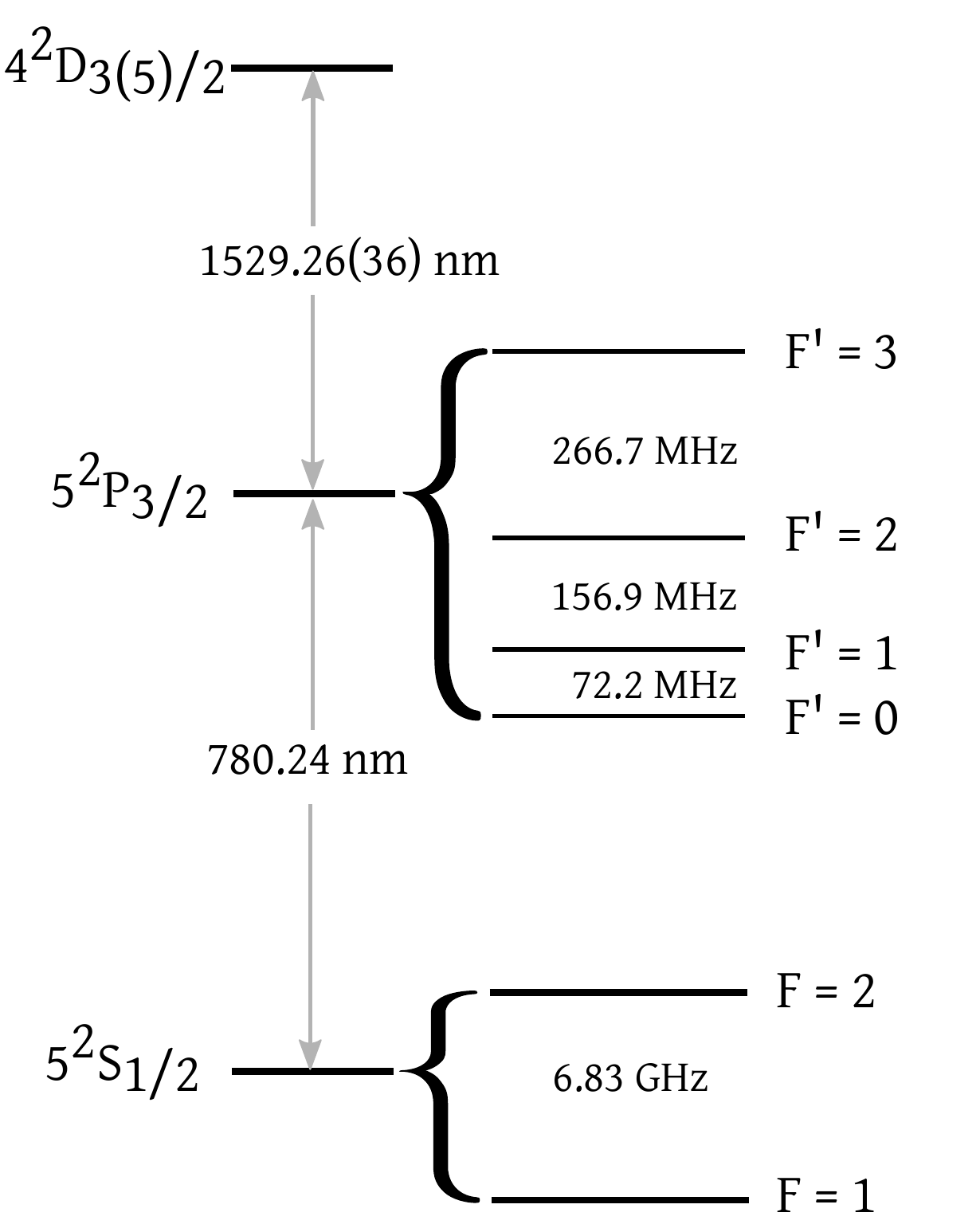}%SBEC_setup.pdf}
\includegraphics[ width=0.5 \columnwidth]{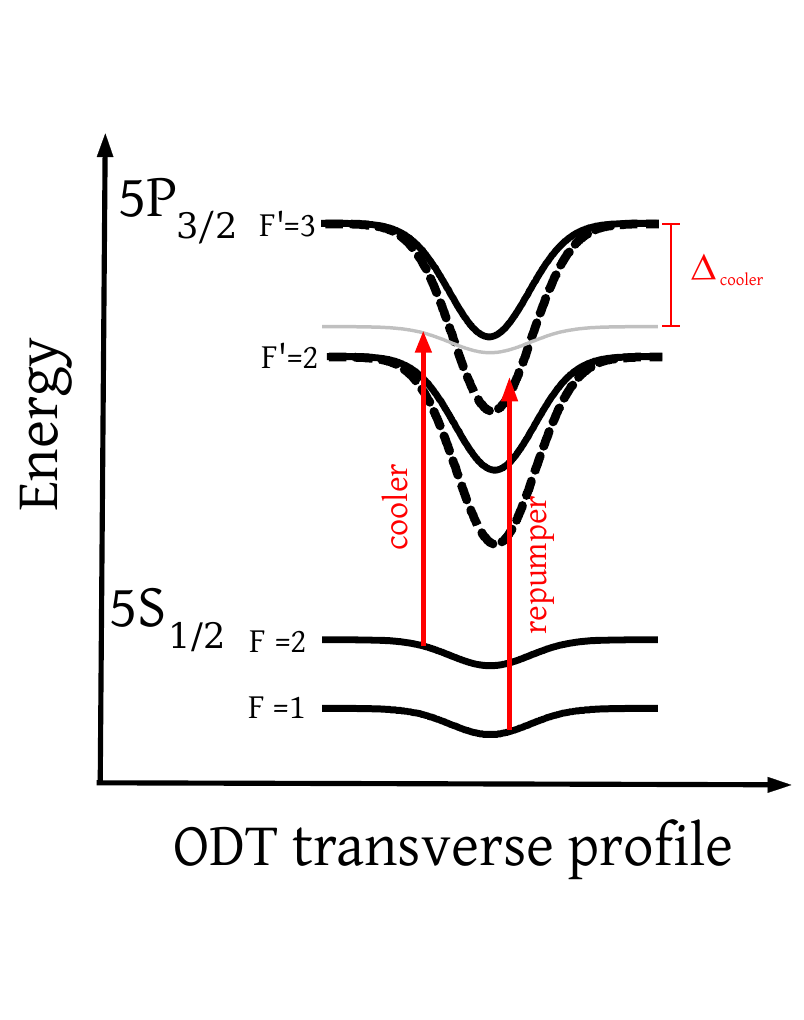}
 \caption{(left) Relevant energy levels of $\rb$. (right) Illustration of differential light-shift  as a function of the position in the optical dipole trap. Detuning of the cooler laser defined as $2 \pi \Delta \equiv \omega_{\rm laser} - (\omega_{F'} - \omega_{F})$ thus positive values of $\Delta$ indicate blue detunings from the  transition. Dashed lines indicate the light-shift at full power of ODT1 with no compensation beam present, whereas continuous lines indicate the light-shift when the trap is partially compensated with the \SI{1529.22}{nm} beam.
 }
 \label{fig:elevels}
\end{figure}

The laser system  is built up around a single  ``master laser'' --  a low noise, narrow linewidth laser that serves as a frequency reference for offset locking of the other lasers.  The master laser is a \SI{1560}{nm} fiber laser (Koheras Adjustik, NKT Photonics) amplified by an erbium-doped fiber amplifier (EDFA) to a maximum power of \SI{3}{W} (Boostik, NKT Photonics), it is frequency-doubled in a periodically-poled LiNbO$_3$ (PPLN) crystal of \SI{50}{mm} length.    The output at \SI{780}{\nano\meter} has a maximum power of \SI{170}{\milli\watt} and Voigt  linewidth of about \SI{4.75+-0.06}{\kilo\hertz} \footnote{The linewidth was estimated from the measurement of the linewidth of the \SI{1560}{nm} laser in a self-heterodyne interferometer with a \SI{0.5}{ms} delay line. The analysis assumes the model proposed in \cite{Mercer} where  the noise is modeled by white noise plus a $1/f$ component, which is  due to thermal fluctuations. The first source of noise gives a Lorentzian character to the linewidth whereas the second one is Gaussian to good approximation. The convolution of both contributions results in a Voigt profile}. The master laser is locked  \SI{80}{MHz} to the blue side of the \transition{F=2}{F'=3} cooling transition (see \autoref{fig:elevels}) using modulation transfer spectroscopy \cite{Natali2015}.
The cooling and repumper lasers  (``slave lasers'') are extended-cavity diode lasers (ECDLs) (Toptica) which are offset locked to the master laser using  an optical phase-locked loop (OPLL), as described in \cite{Appel}, where the ultrafast photodiode is a PIN receiver (PT10GC, Bookham) and the digital phase-frequency-discriminator chip is  an ADF4110 (Analog Devices) in the case of the cooler and an  ADF41020 for the repumper.  The chips are interfaced with a micro-controller that allows us to re-program the loops during the experiment,  thereby tuning the frequency of the slave laser.

In the glass cell, a 3D  magneto-optical trap (MOT) is  formed with a gradient field  of \SI{11.2}{G/cm}, generated by anti-Helmholtz coils mounted around the cell  along the z axis. The bias field is compensated with three pairs of Helmholtz coils in each axis. The six, circularly-polarized beams of the MOT have waists of \SI{1}{cm}  (propagating along $\pm x$ and $\pm z$ directions) or \SI{0.5}{cm}  (propagating along the $\pm y$ direction).  Each beam contains both  cooling and re-pumping light with maximum  intensities of  $\SI{14}{mW/cm^2}$ and $\SI{0.3}{mW/cm^2}$, respectively. The cooler beam is \SI{15}{MHz} red detuned from the \transition{F=2}{F'=3} cooling  transition and the repumper is resonant with the \transition{F=1}{F'=2} transition.  The steady state number of atoms in the MOT is  $10^8$ atoms at $\SI{200}{\micro \kelvin}$.

From the 3D MOT, the atoms are transferred to an optical dipole trap (ODT), formed at different stages of the experiment by up to three ODT beams.  Each beam is linearly polarized, with wavelength \SI{1560}{\nano\meter}  and  Gaussian spatial profile.  ODT1 is  focused to a waist of $\SI{45}{\micro\meter}$ at the center of the MOT, with maximum power of \SI{11}{\watt} and is vertically polarized. ODT2 and ODT3 have waists \SI{65}{\micro m} and horizontal polarization with maximum powers \SI{10}{W} and \SI{7}{W} respectively.  ODT1 and ODT2 propagate along the diagonals in the $y$--$z$ plane, with ODT3 at a $6^\circ$ angle relative to ODT1 (\autoref{fig:setup}).  Acousto-optic modulators are used to shift ODT2 (ODT3) by $-\SI{40}{MHz}$ ($+\SI{40}{MHz}$) relative to ODT1, to avoid spatial interference. In addition, a ``compensation'' beam at \SI{1529.22}{nm} with up to \SI{6}{\milli\watt} of power, mode-matched to ODT1 but of orthogonal linear polarization, can be introduced.

Because the wavelength of the ODT is close to the \SI{1529}{\nano\meter} 3P--4D transition of $\rb$, the ac-Stark shift of the excited 3P state is much stronger than that of the ground state.  The ratio of shifts is about 47.7, which is the ratio of the scalar polarizabilities of the different states \cite{bernon_thesis}. As a result, a large differential light-shift is induced on the D2 line within the ODT position and thus the cooler and repumper lasers have spatially dependent detunings (see \autoref{fig:elevels}). We exploit this fact to form an effective dark MOT similar to that described in \cite{Clement2009}: the atoms at the bottom of the potential are very unlikely to be re-pumped back into the F=2 manifold, and thus they accumulate in F=1, avoiding light-assisted collisions and radiation trapping. 

ODT1 at maximum power creates a differential light shift of \SI{312}{MHz} at beam center, which is larger than the hyperfine splitting between the unshifted \state{F'=2} and \state{F'=3} states. As a result, it is not possible to simultaneously address the cooling transition with red detuning in all spatial regions, as shown in  \autoref{fig:elevels}. To be able to continue the cooling without exciting the \transition{F=2}{F'=2} transition during the loading of the trap we compensate the differential light shift with a \SI{1529.22}{nm} beam, which is blue detuned from the \transition{5P_{3/2}}{4D_{3(5)/2}} transitions.  The light shift induced by the compensation beam in the $\state{5P_{3/2}}$ manifold ranges from \SI{95}{MHz}  to  \SI{109}{MHz} for the different $\state{F'=3, m_F}$ sublevels, according to Floquet-theory estimates, as detailed  in \cite{SimonArxiv}. In the presence of both \SI{1560}{nm} and the \SI{1529.22}{nm} beams, the differential light shift at the bottom of the trap  is reduced to \SI{210}{MHz}.

In the partially compensated dipole trap  a molasses phase is started: the magnetic field gradient is suddenly switched off, the cooling laser is further detuned to \SI{255}{MHz}  to the red of the unshifted \transition{F=2}{F'=3} transition, and thus \SI{45}{MHz} red-detuned at the bottom of the trap. The power of the  repumper is lowered by a factor of 2.5 without changing the frequency. This phase lasts \SI{500}{ms},   limited by the lifetime of the cold atom reservoir. This strategy allows us to load up to \SI{7e6}{atoms} in $\state{F=1}$ at $\SI{50}{\micro K}$ into the dipole trap. Using the compensation beam therefore improves the maximum number of atoms loaded by a factor of three respect to the non-compensating strategy of the ODT1 at full power, and by a factor of two loading ODT1  at a lower power for which the differential light shift does not exceed the excited-state hyperfine splitting. 

After the ODT is loaded,  the cooler and repumper beams are switched off and the power of the  compensation beam adiabatically lowered. At the same time, a magnetic field of magnitude $B_z=\SI{1}{G}$ is applied along the $z$ axis. At this field, the atoms are optically pumped into the $\state{F=1,m_F=+1}$ state using a beam  (OP) resonant with the \transition{F=1}{F'=1} transition  and propagating along the $z$ axis  with $\sigma^+$ polarization. We achieve $90 \%$ efficiency of pumping as confirmed by  Stern-Gerlach imaging along the quantization axis defined by the magnetic field. To avoid the effects of the spatially-dependent differential light shift on the atoms distributed in the trap, the optical pumping is done with three \SI{20}{\micro s} OP pulses  during which the ODT1 is switched off. The pulses are separated by  \SI{10}{ms} intervals to allow the atoms to redistribute in the trap and avoid shadowing effects.

 Following optical pumping the atoms are allowed to thermalize for \SI{500}{\milli\second}, after which the cloud is compressed in the longitudinal direction of ODT1 using ODT3. ODT3  boosts the collision rate without reducing the large collection volume. Forced all-optical evaporation in this two-beam trap is possible and efficient down to \SI{2}{\micro K}. At that point the longitudinal frequency becomes insufficient to reach higher phase space densities. We employ an extra beam, ODT2, to provide extra compression at the end of the evaporation. 

 The evaporation sequence is as follows: starting with all the three beams at full power, we perform forced all-optical evaporation  for  \SI{4}{s}, after which the system crosses the critical temperature $T_c$  with about $10^5$ atoms. The power of ODT2 is then increased for an additional 800 ms to compress the atoms, resulting in the formation of  a pure condensate with typically $\SI{4e4}{atoms}$. The relative populations do not change during the evaporation, as discussed in \autoref{sec:relaxation} below.

\newcommand{\RTF}{R_{\rm TF}}
\newcommand{\bRTF}{\bar{R}_{\rm TF}}

\section{Single spin domain \label{sec:spat}}

 A natural measure of the spatial extent of the condensate is the Thomas-Fermi radius.
The mean  can be expressed in terms of fundamental constants, the number of condensed atoms $N$ and the mean oscillation frequency of the harmonic trap $\bar{\omega}$:
 
\begin{equation}
 \bRTF= \left( \frac{15 N a_0 \hbar^2}{M^2\bar{\omega}^2}\right)^{1/5}
 \label{Rtf}
\end{equation}
were $a_0=\SI{5.38e-9}{m}$ is the scattering length and  $M=\SI{1.44e-25}{kg}$ the mass of one \rb atom. 
%,   $\bar\omega = \sqrt[3]{\omega_1 \omega_2 \omega_3}$, and $\omega_i$ is the trap frequency along the $i$th principal axis.  
The number of atoms is measured with time-of-flight  absorption imaging for dense clouds \cite{Reinaudi2007}.  

\begin{figure}[h!]
\includegraphics[width=\columnwidth ]{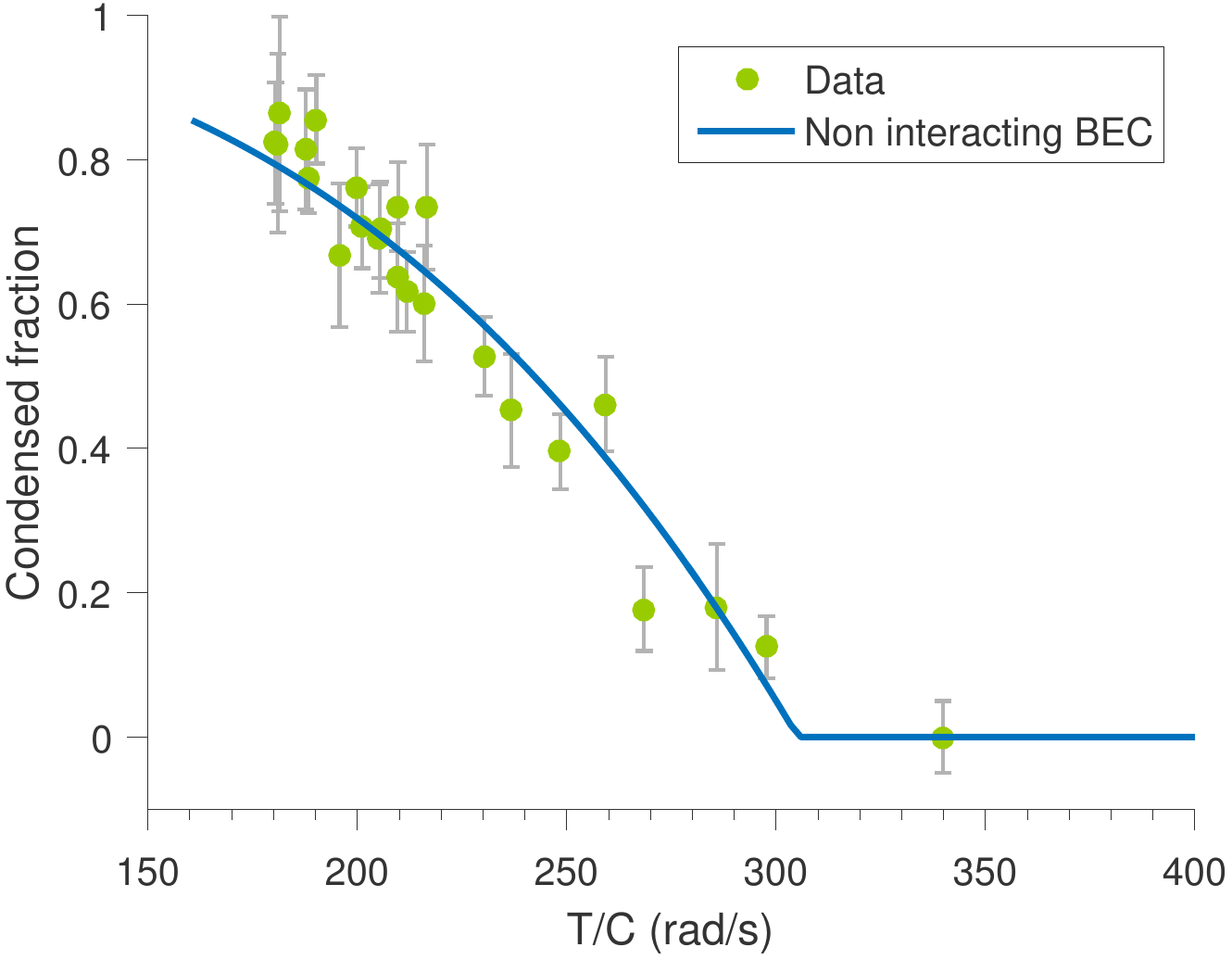}
\caption{Condensed fraction as a function of $T/C$ where the temperature $T=\SI{54}{nK}$ is constant and $C$ is a function of the number of atoms, \Ntot.  Since only  \Ntot varies  the fit model to $c_F=\max \{1-[T/(C\bar{\omega}_f)]^3,0\}$ where $\bar{\omega}_f$ is the only free parameter, yielding to $\bar{\omega}=2\pi \cdot\SI{50+-5}{Hz}$. }
\label{cf}
\end{figure}

The relative low atom number prevented direct measurement of the trap frequencies by parametric excitation. Instead we estimate $\bar{\omega}$ from the observable condensate fraction as follows:  for $N$ condensed atoms of a total $N_{\rm tot}$ atoms, the condensed fraction $c_F=N/N_{tot}$  for non-interacting bosons obeys the relation: $c_F=\max[1-(T/T_c)^3,0]$, where  the critical temperature in a harmonic trap is  \cite{Pethick}:
 
\begin{equation}
T_c= \frac{\hbar \bar{\omega}}{k_B} \left[\frac{N_{tot}}{\zeta (3)}\right]^{1/3} 
\label{eq:tc}
\end{equation}
$k_B$ is the Boltzmann constant and $\zeta(3)\approx 1.202$. Time-of-flight absorption imaging and a bi-modal fit give direct access to $N_{\rm tot}$, $N$ (and thus $c_F$), as well as to $T$, which is  found from the width of the thermal component. For fixed beam geometry and power, the trap frequencies are constant. The temperature is set by the potential depth at  $T=\SI{54+-13}{nK}$, independent of $N_{\rm tot}$. We can thus vary the critical temperature by changing only the number of atoms, through the duration of the ODT loading step. \autoref{cf} shows the condensed fraction as a function of $T/C$, where $C  \equiv T_c/\bar{\omega}$. We fit the expected  scaling of $c_F$   where $\bar{\omega}$ is the only free parameter, to find $\bar{\omega}=2\pi \cdot\SI{50+-5}{Hz}$. 

 The mean Thomas-Fermi radius as given by \autoref{Rtf} with  $N=\SI{4e4}{}$ atoms is then: $\bRTF=\SI{7.0+-0.5}{\micro m}$. From the geometry and power of the ODT beams the shape of the condensate is expected to be a spheroid with ${\RTF}_ 1={\RTF}_ 2$ and anisotropy factor given by $l(t)\equiv {\RTF}_ 1(t)/{\RTF}_ 3(t)$. We measured the size of the condensate after \SI{30}{ms} of time-of-flight and found the final anisotropy to be $l(t=\SI{30}{ms})=\SI{1.63+-0.03}{}$. Using the expansion model \cite{Castin1996}, we have estimated the initial anisotropy to be $l(t=0)=0.84$.  The assumption of a spherical geometry is thus not expected to introduce large errors in what follows.  

 To gain some intuition about the spin-dependent contribution to the magnetization distribution the $\RTF$ radius is compared to the spin healing length, which is defined as $\xi_s=\hbar/\sqrt{2 M |c_1| n}$, where $c_1=\SI{-2.39e-54}{Jm^3}$ characterizes the spin-dependent contact interaction, and $n$ is the density \cite{Stamper2001, Kawaguchi}.   In our experimental conditions, $n =\SI{2.7e13}{\per\centi\meter\cubed}$   and  $\xi_s= \SI{7.7}{\micro m}$, while the density healing length is $\xi_n=\hbar/\sqrt{2 M c_0 n}=\SI{0.5}{\micro m}$. %and the dipolar healing length is $\xi_d=\hbar/\sqrt{2 M c_{dd} n}=\SI{25}{\micro m}$. 
 Although density variations are possible since $\xi _n \ll \bRTF$, spin variations in space are unlikely to occur since $\xi _s \sim \bRTF$ and therefore is energetically unfavorable for the condensate to split into different spin domains,which  would allow loss of coherence as the domains dephase relative to each other due to for example field gradients.  In this regime the SMA can capture all the physics of the spinor condensate.

\section{Excitation of spin oscillation and free oscillation \label{sec:oscB}}

Within the SMA the order parameter can be written as $\mathbf{\hat{\Psi}}(\vec{r})=\varphi(\vec{r})\mathbf{\hat{\Psi}}$, where $\varphi(\vec{r})$ defines the spatial mode which is common between  all the spin states. We write the spinor as $\mathbf{\hat{\Psi}} \equiv \spinor{\hat{\psi}_{+1}, \hat{\psi}_{0}, \hat{\psi}_{-1}}$, where $\hat{\psi}_{m_F}$ are the complex amplitudes describing the magnetic sublevels  \cite{LawPRL1998}. 
The atoms condense in $\state{F=1,m_F=+1}$ so initially $\mathbf{\hat{\Psi}} = \spinor{1,0,0}$. 

From this initial state, we lower $B_z$ to zero while increasing the field along  $x$ to  a maximum value $B_x$. This is done slowly compared to the Larmor frequency defined by the field amplitude so the spins adiabatically follow the  field and finish pointing along $x$. To rotate the spins to the $y$--$z$ plane and form the state $\frac{1}{2}\spinor{1, \sqrt{2},1}$,  a $\pi/2$-RF pulse along $y$ is applied.  This starts Larmor precession dynamics around the $x$ axis with a Larmor frequency  $\omega_L = \gamma B_x$, where $\gamma = 2 \pi (\SI{-0.7}{MHz/G})$ is the gyromagnetic ratio of $\rb$ in the $F=1$ manifold.  

While the dominant effect on the spin dynamics is the linear Zeeman shift, which produces a Larmor precession about $x$, the quadratic Zeeman shift also plays a significant role, and manifests as a modulation of the Larmor precession. As described below, the $z$ projection of the spin evolves as $\langle {f}_z\rangle(t) = \cos\Theta_L \cos\Theta_Q$ where $\Theta_{L,Q} \equiv \int dt \, \omega_{L,Q}(t)$ and  $\omega _Q= \hbar \gamma ^2 B^2/\Delta E_{\rm hf}$, with $\Delta E_{\rm hf}= 2 \pi \hbar \times \SI{6.83}{GHz}$ being the hyperfine splitting \cite{StamperRev}.

\newcommand{\hj}{\hat{\jmath}}
\newcommand{\hf}{\hat{f}}
\newcommand{\hJ}{\hat{J}}
\newcommand{\hF}{\hat{F}}
\newcommand{\feq}{\stackrel{F=1}{\longrightarrow}}

A single atom exposed to a magnetic field $B_x(t)$ along the $x$ direction experiences the spin Hamiltonian 
\begin{eqnarray}
\label{eq:1AtomHamiltonian}
\hat{H} &=& \frac{g_J \mu_B}{4} B_x \opf{x} + \frac{g_J^2 \mu_B^2}{16 \Delta E_{\rm hf}} B_x^2 \opf{x}^2 + O(B_x^3) \nonumber \\
& \equiv & \hbar \omega_L \opf{x} + \hbar \omega_Q \opf{x}^2 + O(B_x^3)
\label{eq:OneAtomSpinHam}
\end{eqnarray}
where  $\opf{\alpha}$ denotes the spin-1 Pauli matrix for component $\alpha \in \{x,y,z\}$ \cite{Colangelo2013}.%, which obeys the commutation relations $[\opf{x}, \opf{y}] = i \opf{z}$ and cyclic permutations. 
We note that $B_x$ and thus $\omega_L, \omega_Q$ may be time-dependent. 
Due to the $\opf{x}^2$ term the dynamics induced by the Hamiltonian  involves both, the vector ``spin orientation'' components $\opf{x},\opf{y},\opf{z}$, and the rank-two tensor ``spin-alignment'' components: 

\begin{eqnarray}
%j_x &\equiv&  f_x^2- f_y^2 \\
\opj{zx} &\equiv& \opf{z}\opf{x}+\opf{x}\opf{z} ,\\%j_k
\opj{xy} &\equiv&\opf{x}\opf{y}+ \opf{y}\opf{x}. %j_y
%j_l &\equiv& f_y f_z+ f_z  f_y\\
%j_m&\equiv& (2 f_z^2- f_x^2- f_y^2)/\sqrt{3}
\end{eqnarray}
We find the single-atom equations of motion
\begin{eqnarray}
\frac{d}{dt}  \left( \begin{array}{c}
f_x \\ f_y \\ f_z \\ j_{zx} \\ j_{xy}
\end{array} \right) &=&  \left( \begin{array}{ccccc}
0 & 0 & 0 & 0 & 0 \\
0 & 0 & -\omega_L & - \omega_Q & 0   \\
0 & \omega_L & 0  & 0  &  \omega_Q \\
0 & \omega_Q  & 0  & 0 &  \omega_L  \\
0 & 0 & -\omega_Q &   -\omega_L  & 0 
\end{array} \right) \left( \begin{array}{c}
f_x \\ f_y \\ f_z \\ j_{zx} \\ j_{xy}
\end{array} \right)
\end{eqnarray}
which describe a pair oscillators, $f_y$--$f_z$ and $j_{zx}$--$j_{xy}$, each with oscillation frequency $\omega_L$, and mutual coupling frequency $\omega_Q$.  
%For the components $f_x$ and $f_z$, the general solution is 
%\begin{eqnarray}
%f_x(t) &=&  f_x(0) \\
%f_z(t) & = & 
%\left[ f_z(0) \cos \Theta_L +  f_y(0) \sin \Theta_L \right] \cos\Theta_Q
%\nonumber \\ & & +
%\left[ -j_{zx}(0) \sin \Theta_L +  j_{xy}(0) \cos \Theta_L \right] \sin\Theta_Q  \,\,\,\,\\
%\Theta_{L,Q} &\equiv &  \int dt\, \omega_{L,Q} (t) 
%%\Theta_Q &\equiv &  \int dt\, \omega_Q(t) 
%\end{eqnarray}
%
%in the following way:
% \begin{align}
% \opf{x}(t)=&\opf{x}(0), \\
% \opf{y}(t)=&\cos\Theta_Q[\cos\Theta_L\opf{y}(0) -\sin\Theta_L\opf{z}(0)]  \\
% &-\sin\Theta_Q[\cos\Theta_L\opj{zx}(0) +\sin\Theta_L\opj{xy}(0)] , \nonumber\\
% \opf{z}(t)=&\cos\Theta_Q[\cos\Theta_L\opf{z}(0) +\sin\Theta_L\opf{y}(0)] \\
% &+\sin\Theta_Q[\cos\Theta_L\opj{xy}(0) -\sin\Theta_L\opj{zx}(0)] , \nonumber\\
% \opj{xy}(t)=&\cos\Theta_Q[\cos\Theta_L\opj{xy}(0) -\sin\Theta_L\opj{zx}(0)] \\
% &-\sin\Theta_Q[\cos\Theta_L\opf{z}(0) +\sin\Theta_L\opf{y}(0)], \nonumber\\
% \opj{zx}(t)=&\cos\Theta_Q[\cos\Theta_L\opj{zx}(0) +\sin\Theta_L\opj{xy}(0)]  \\
% &+\sin\Theta_Q[\cos\Theta_L\opf{y}(0) -\sin\Theta_L\opf{z}(0)] , \nonumber \\
% \Theta_{L,Q} \equiv &  \int dt\, \omega_{L,Q} (t).
% \end{align}
%which describe a pair \btext{of coupled} oscillators, $\opf{y}$--$\opf{z}$ and $\opj{zx}$--$\opj{xy}$, each with oscillation frequency $\omega_L$, and mutual coupling frequency $\omega_Q$.

The collective spin of $N$ condensed atoms is  ${\bf F} \equiv \sum_{n=1}^{N} {\bf f}^{(n)}$, where the superscript indicates the $n$th atom The dynamics of the collective spin then obeys 
\begin{eqnarray}
\opF{i}(t) &=& N(t) \opf{i}(t) 
%\opF{z}(t) &=& \Ntot(t) \opf{z}(t)
\end{eqnarray}
which describes the coherent oscillation  of a macroscopic spin, and its decay caused by the loss of atoms. This is studied further  in \autoref{sec:relaxation}.

\section{Non-destructive probing of the spin polarization  \label{sec:farprob}}

\begin{figure}[ht!]
     \includegraphics[width = \columnwidth ]{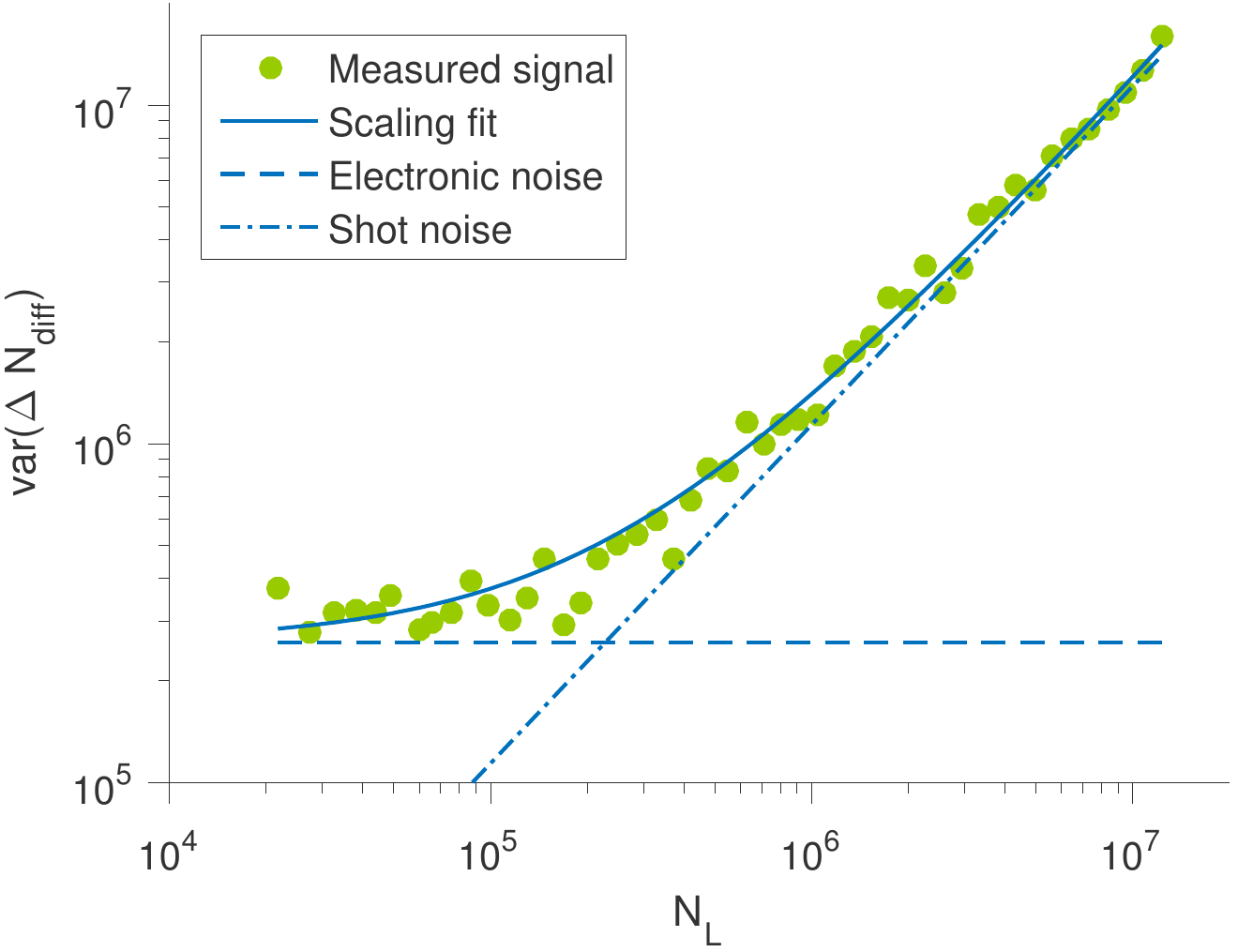}
          \caption{Variance of the output signal of the differential photodetector as a function of total input number of photons ($\NL$). The fit assumes the form $\var{\Delta N_{\rm diff}} = V_e + N_L$ to find the electronic noise floor $V_e = \num{2.6+-0.5e5}$.}
      \label{fig:detector}
\end{figure}

We perform non-destructive Faraday rotation measurements of the spin state of the atoms, exploiting the spin-dependent interaction of a linearly polarized off-resonant beam with the vectorial component of the atomic polarizability, as described in \cite{Koschorreck2010}. The interaction with the atoms causes the linear polarization of this beam to rotate and therefore to acquire a diagonal component, which is detected using a shot-noise limited polarimeter based on the  differential photodetector (DFD) described in \cite{Martin2016}.  We have demonstrated the differential photodetector is shot-noise limited for \SI{}{\micro\second} pulses with \num{2e5} to \SI{e7}{photons}, having an electronic noise floor equivalent to the shot noise of a pulse with \SI{2.6+-0.5e5}{photons}, as shown in \autoref{fig:detector}.

The Faraday beam is  red detuned \SI{276}{MHz} from the \transition{F=1}{F'=0} transition  and has linear polarization of \ang{54.7}  with respect to the bias field along $x$. At this ``magic angle''  the tensorial AC stark shift averages to zero over one precession cycle  \cite{Smith2004}, enabling continuous probing without conversion of spin alignment to spin orientation. 

\begin{figure}[ht]
     \includegraphics[width = \columnwidth ]{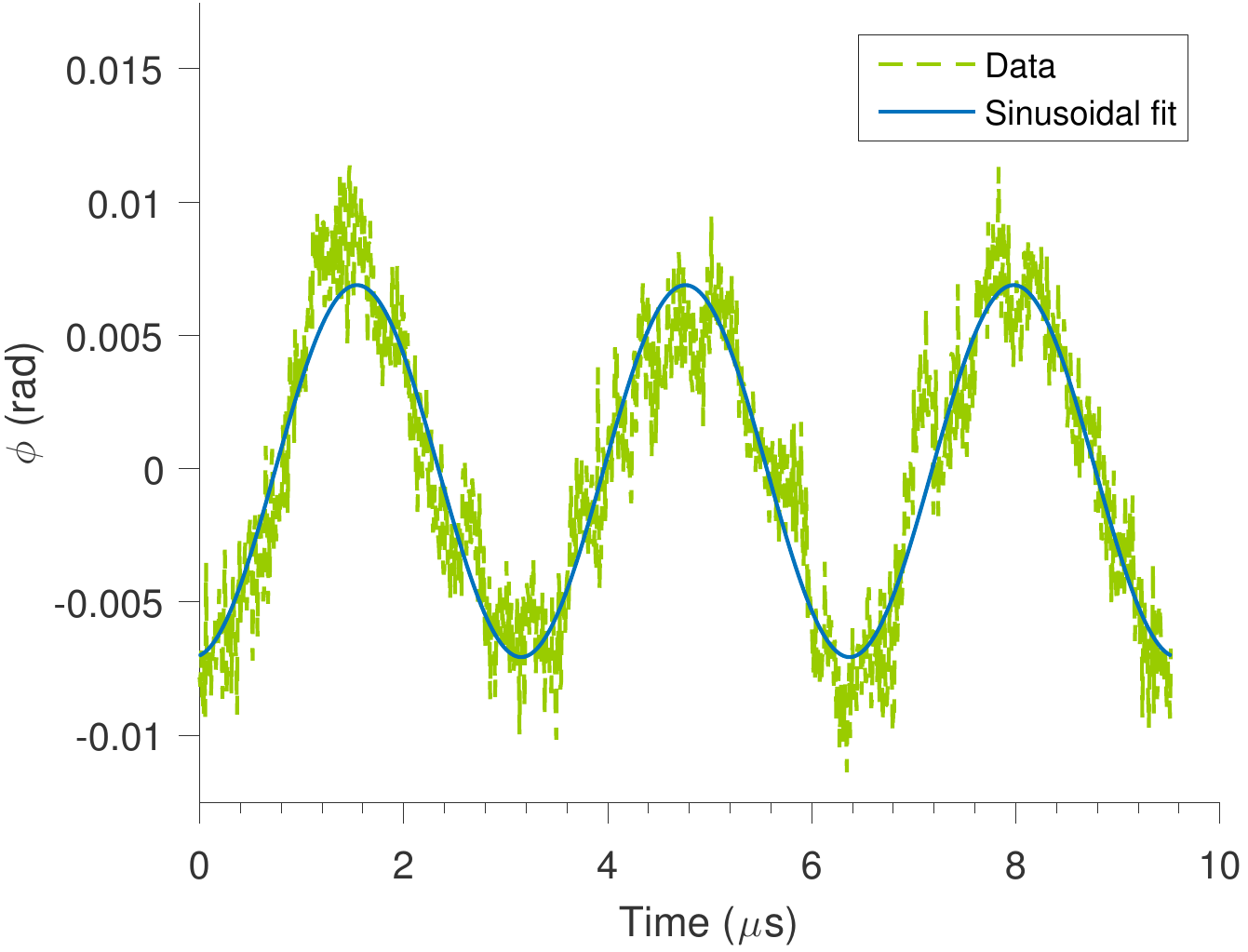}
     \caption{Example of Faraday rotation angle along the probing direction z. In this example, the probing pulse contains \SI{5e6}{photons} and the magnetic field is such that the Larmor frequency is $\omega_L= 2 \pi \cdot \SI{300}{kHz}$. The coupling constant is such that $\phi(t)=(\SI{e-7}{rad/spin})F_z(t)$.}
     \label{fig:exampleFar}
\end{figure}

The Faraday probing  is typically performed with $\tau_p=\SI{10}{\micro s}$ pulses containing $\NL=5\times 10^6$ photons. To compensate probe power fluctuations, $N_L$ is measured by splitting a fraction of the power to  an auxiliary PD and transimpedance amplifier before entering the chamber. An example of the Faraday rotation signal is shown in \autoref{fig:exampleFar}. The polarization rotation angle in the Poincar\'e sphere $\phi = G_1 F_z$ is  proportional to the collective spin component $F_z$, and with a coupling $G_1$ that depends on the overlap of the beam and the atomic cloud. For this reason, the probing beam is focused at the atoms position with a waist of $\SI{18}{\micro m}$.  For the pure condensate, we observe $G_1=\SI{e-7}{rad/spin}$, by measuring the rotation angle caused by a fully polarized cloud with a known number of spins calibrated with absorption imaging.  

 The known shot-noise scaling of the optical angle  allows us to estimate for  one pulse with $\NL=\SI{5e6}{photons}$ an optical angle noise of $d\phi=1/\sqrt{\NL} = \SI{0.4}{mrad}$, which corresponds to an inferred  noise in the spin state $\sqrt{\var{F_z}}=(G_1\sqrt{\NL})^{-1}\simeq \SI{4e3}{spins}$. %(\num{e7}\times \num{.4e-3})$.
The noise is larger than the projection noise inherent to the atomic state, which is given by $\sqrt{\var{F_z}}=\sqrt{N}=\SI{2e2}{spins}$.  The interaction with each pulse does not cause atoms to be lost from the trap, but it kicks atoms out from the condensate reducing the condensed fraction by $25\%$. Reaching low damage projection noise limited measurements requires improvement to the coupling factor $G_1$ rather than using more photons.  

\section{Spin coherence properties of the spinor BEC \label{sec:relaxation}}

In other spin systems, e.g. liquid-state magnetic resonance \cite{Bloembergen1948}, it is common to distinguish a relaxation rate $1/\Tone$ for the longitudinal spin component, i.e., the one parallel to the field, and distinct transverse relaxation rates $1/\Ttwo$  due to only homogeneous effects (relevant for spin-echo experiments), and $1/\Ttwo^*$ for relaxation due to both homogeneous and inhomogeneous effects.  Here, in contrast, the single-mode condition implies a single spin state $\mathbf{\hat{\Psi}}(t)$ for the entire condensate, enforcing full coherence. We thus expect $\Ttwo = \Ttwo^* = \Tone$, with $\Tone$ limited only by loss of atoms from the condensate, to which we assign a rate $1/\Tzero$.

\textit{Atom losses}: Atom losses are caused by collisions with the background gas and by three-body collisions. The former knock condensate atoms out of the trap, or less frequently into the thermal cloud. The latter are strongly exothermic and result in loss of all three atoms. The atom number in the condensate evolves as 
\begin{equation}
\dot{N}=-N/\tau - K_3 \int d^3x \, n^3(x,t)
\end{equation}
where the first term describes loss from background collisions and the second from three-body losses \cite{Burt1997, Soding1999}. For condensed atoms of  \rb, $K_3 \simeq \SI{6e-30}{\centi\meter^6\per\second}$ so that at our densities of $n \simeq \SI{3e13}{\per\centi\meter\cubed}$, the three-body loss rate is of order $K_3 n^2 N \simeq N/\SI{200}{\second}$.  In contrast, the observed number decay, measured by absorption imaging, is much faster, and well described by one-body losses with $\Tone = \tau = \SI{7.7 +- 0.4}{s}$. The three-body loss can thus be neglected in these conditions. 

\textit{Longitudinal spin relaxation}: Under the Hamiltonian of \autoref{eq:OneAtomSpinHam} above, both $f_x$ and the magnetic quantum number $m_F$ (in the $x$-basis) are constants of the motion, even for fluctuating $B_x(t)$.  This we confirm using Stern-Gerlach imaging to measure the population in the different magnetic sublevels as a function of hold time: a condensate is prepared in the $\spinor{1,0,0}$ state and held in the dipole trap during a time $t_e$, after which the atoms are released from the trap. During the time of flight a gradient field of $\sim \SI{20}{G/cm}$ is applied for \SI{10}{ms} to spatially separate the different spin components, before performing absorption imaging. The relative populations of the different spin states remain unchanged as a function of $t_e$, to within measurement precision. 

We note that orthogonal AC magnetic fields at a frequency close to $\omega_L$ could resonantly excite transitions among $m_F$ levels. The influence of such fields has limited the observation of spin dynamics in other experiments \cite{Chang2004}. In our experiment this effect becomes evident only at bias fields below \SI{100}{mG}.
   
   \textit{Transverse spin relaxation}:  Very long coherence times require very stable bias fields to  estimate $\Ttwo^*$ by  direct measurement of $\langle F_z(t) \rangle$.  A change in the magnetic field between repetitions of the experiment  generates a  $F_z(t)$ with a different phase  $\Theta_L$, and the average  shows a relaxation-like behaviour $\langle F_z(t) \rangle \propto \exp[-\tau/\Ttwom]$, where $\Ttwom$ is a relaxation time related to the shot-to-shot variation of the field, and only weakly related to the processes described by $\Ttwo^*$.  Using a fluxgate sensor, we measured the spectrum of environmental magnetic noise near the atoms.  Simulating the spin dynamics for such variations of the magnetic field we estimate $\Ttwom = \SI{1.3}{ms}$. 

To accurately measure  $\Ttwo^*$ in such circumstances, we take advantage of the non-destructive nature of the Faraday rotation probing, which allows us to probe several  Larmor cycles of $F_z(t)$ during a single run and extract the amplitude via a sinusoidal fit.  Because the amplitude does not depend on $\Theta_L$, this allows meaningful averaging in spite of the shot-to-shot fluctuations. The quadratic phase $\Theta_Q$ also varies shot-to-shot, but on a time-scale about four orders of magnitude longer than does $\Theta_L$, implying that the average amplitude  will show effects of dephasing on the $\SI{10}{s}$ time scale.

We prepare the state $\frac{1}{2}\spinor{1, \sqrt{2},1}$ and allow the atoms precess around the bias field $B_x=\SI{0.29}{G}$ for time $t_e$, before $F_z(t)$ is measured  for \SI{10}{\micro s} allowing several Larmor cycles to be resolved by fitting a sinusoidal function as in  \autoref{fig:exampleFar}. We perform measurements at different values of $t_e $ ranging from $0$ to \SI{1}{s}, each one on a new preparation of the state. The Larmor frequency is always  $(\omega_L=2 \pi \cdot \SI{200}{kHz})$. To separate the relaxation and decoherence signature from the atomic losses, we normalized the signal by the number of atoms measured by absorption imaging at the end of each repetition.

\begin{figure}[ht]
      \includegraphics[width = \columnwidth]{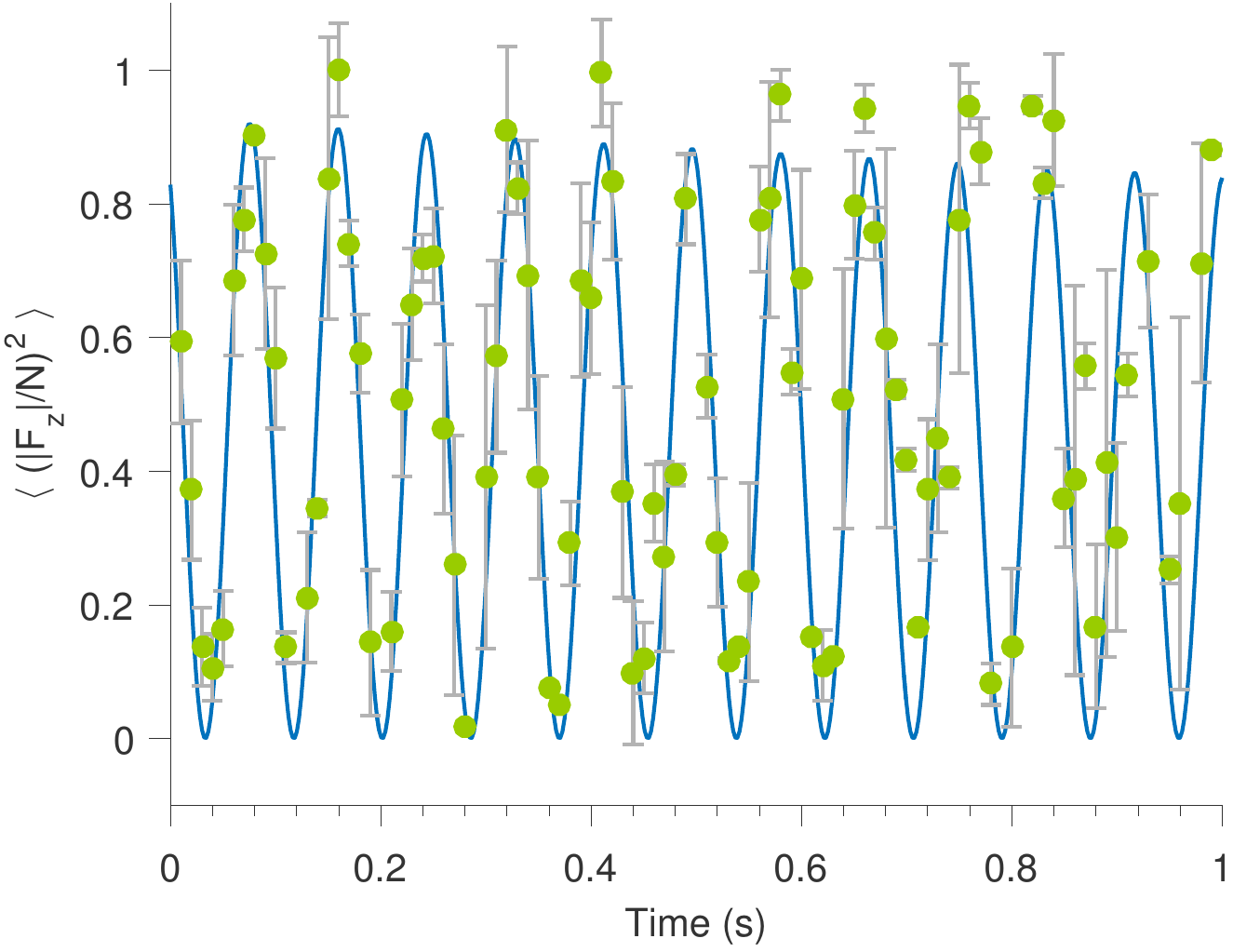}
 \caption{Average over different repetitions of the normalized z spin projection square vs evolution time (s). The signals are obtained using  Faraday rotation measurements of a fully polarized condensate. The spins rotate around $x$ at Larmor precession frequency $\omega_L = 2 \pi \cdot  \SI{200}{kHz}$ with a quadratic component oscillating at $\omega_Q= 2 \pi \cdot \SI{6}{Hz}$. We observe negligible decay of the normalized magnetization, which implies $\Ttwo ^*\simeq \Tzero$. }%data in 2016/11/21/12_00_20. The field is lowered such that nu_L=200kHz
 \label{fig:T2*}
\end{figure}

In \autoref{fig:T2*} we plot the mean square amplitude   of the sinusoidal fits for different repetitions, as a function of $t_e$.  To this data we fit  a function $ A \cos^2(\omega_Qt)\exp{(-2t/\mathcal{T})}$, from which we find  $\omega_Q=2\pi \cdot\SI{5.95+-0.01}{Hz}$ and $\mathcal{T} \gg \SI{1}{\second}$, much longer than the observation time, which implies $\mean{F_z}$ is only limited by the atom losses and therefore $\Ttwo^*\simeq \Tzero$. As can be appreciated in the figure, the full visibility is always recovered even at observation times as long as \SI{1}{s}. 

With these results we confirm $\Ttwo^*, \Ttwo , \Tone \gg \SI{1}{\second}$. That is, we observe no relaxation or decoherence mechanism degrading the coherence of the state in the observed time scales. The result is consistent with the expectation of spin relaxation due solely to atom losses, which in our typical vacuum conditions is limited  by one-body losses $\Ttwo^*\simeq \Ttwo \simeq \Tone \simeq \Tzero = \SI{7.7 +- 0.4}{s}$.

\section{Conclusions}

This work reports on the construction of a minimalist system capable of creating single-mode spinor Bose-Einstein condensates of \rb atoms in the ferromagnetic $F=1$ hyperfine state.  The loading of the dipole trap involves a novel  technique where the differential light shift induced  by the dipole trap is exploited to create an effective dark-MOT. 
Based on measurements of the spatial size, densities and coherence, we demonstrate the spinor condensate is created in a single spin domain, because there are not different domains the collective spin does not dephase in the time scales limited by the lifetime of the condensate. We have demonstrated one second of coherence of the collective spin state and inferred several seconds  of coherence via non-destructive probing of the spin state based on Faraday rotation measurements. 
The noise of the probing is close to the atomic projection noise inherent to the spin state.  The long coherence together with the small size, situate the system  in a very promising position for applications including coherent sensing of magnetic fields. 

\vfill
%\bibliographystyle{plainnat}
%\bibliography{../biblio/SD-SBEC}
\bibliography{SD-SBEC}

\end{document}